\documentclass[useAMS,usenatbib,aas_macros]{mnras}
\usepackage{graphicx}
\usepackage{epstopdf}
\usepackage{bm}
\usepackage{times}
\usepackage{amsmath,amssymb}
\usepackage{slashed}
\usepackage{color}
\usepackage{natbib}
\usepackage{hyperref}
\usepackage[]{appendix}
\usepackage{multirow}
\usepackage{tabularx}

\begin{document}

\title[Deep XMM Observations of Draco rule out a dark matter decay origin for the 3.5 keV line]{Deep XMM Observations of Draco rule out at the 99\% Confidence Level a Dark Matter Decay Origin for the 3.5 keV Line}

\author[T. Jeltema and S. Profumo]{Tesla Jeltema$^{1}$\thanks{tesla@ucsc.edu} and Stefano Profumo$^{1}$\thanks{profumo@ucsc.edu}\\
$^{1}$Department of Physics and Santa Cruz Institute for Particle Physics
University of California, Santa Cruz, CA 95064, USA
}


\maketitle

\begin{abstract}
We searched for an X-ray line at energies around 3.5 keV in deep, $\sim 1.6$ Msec XMM-Newton observations of the dwarf spheroidal galaxy Draco. No line was found in either the MOS or the PN detectors.  The data in this energy range are completely consistent with a single, unfolded power law modeling the particle background, which dominates at these energies, plus instrumental lines; the addition of a $\sim 3.5$ keV line feature gives no improvement to the fit.  The corresponding upper limit on the line flux rules out a dark matter decay origin for the 3.5 keV line found in observations of clusters of galaxies and in the Galactic Center at greater than 99\% C.L..  
\end{abstract}


\maketitle

\section{Introduction}
The detection of a line with an energy between 3.50 -- 3.57 keV (hereafter indicated as ``the 3.5 keV line'' for brevity) in the X-ray data from individual and stacked observations of clusters of galaxies \citep{Bulbul:2014sua}, from the Galactic center \citep{Jeltema:2014qfa} and, tentatively, from M31 \citep{Boyarsky:2014jta} \citep[see however][]{Jeltema:2014qfa, reply} has triggered widespread interest: the line might be associated with a two-body radiative decay including one photon of a dark matter particle with a mass of around 7 keV and a lifetime of about $6-8\times10^{27}$ sec. Such a particle has a natural theoretical counterpart in sterile neutrino models, a class of dark matter candidates whose motivation goes beyond that of explaining the missing non-baryonic matter in the universe \citep[see e.g.][for a review]{Boyarsky:2009ix}.

\cite{Jeltema:2014qfa} pointed out early on that atomic de-excitation lines from He-like Potassium ions (K XVIII) are a plausible counterpart to the 3.5 keV line both in clusters of galaxies and in the Milky Way. This possibility was initially discarded by \cite{Bulbul:2014sua} based on estimates of the required K abundance that relied on photospheric K solar abundances, and on multi-temperature models biased towards high temperatures. The latter, as demonstrated in \cite{reply}, artificially suppress the brightness of the K XVIII de-excitation lines by up to more than one order of magnitude. Additionally, coronal K abundances are larger by about one order of magnitude than photospheric K solar abundances, as recently pointed out by \cite{Phillips:2015wla}. As a result, the case for K XVIII as the culprit for the 3.5 keV line appears at present quite plausible. 

Additional circumstantial evidence against a dark matter decay origin for the 3.5 keV line has also emerged. \cite{Malyshev:2014xqa} searched for the line in stacked, archival XMM observations of dwarf spheroidal galaxies, reporting a null result that highly constrained a dark matter decay origin for the line. \cite{Anderson:2014tza} analyzed stacked observations of galaxies and galaxy groups, systems where the thermal emission would be too faint to produce a detectable line from e.g. K XVIII, and also failed to find any evidence for a 3.5 keV line. \cite{Urban:2014yda} studied Suzaku data from X-ray-bright clusters, confirming that the 3.5 keV signal could naturally be ascribed to K, and questioning the compatibility of the line morphology with the dark matter decay hypothesis.  Finally, \cite{Carlson:2014lla} studied in detail the morphology of the 3.5 keV emission from the Perseus cluster of galaxies and from the Galactic center, finding a notable correlation with the morphology of  bright elemental emission lines, and excluding a dark matter decay origin even for cored Galactic dark matter density profiles. A recent study of charge exchange processes indicates that an additional possibility is that the 3.5 keV line originates from a set of high-$n$ S XVI transitions (populated by charge transfer between bare sulfur ions and neutral hydrogen) to the ground state \citep{Gu:2015gqm}.

\begin{figure*}
\begin{centering}
	\includegraphics[width=0.93\columnwidth]{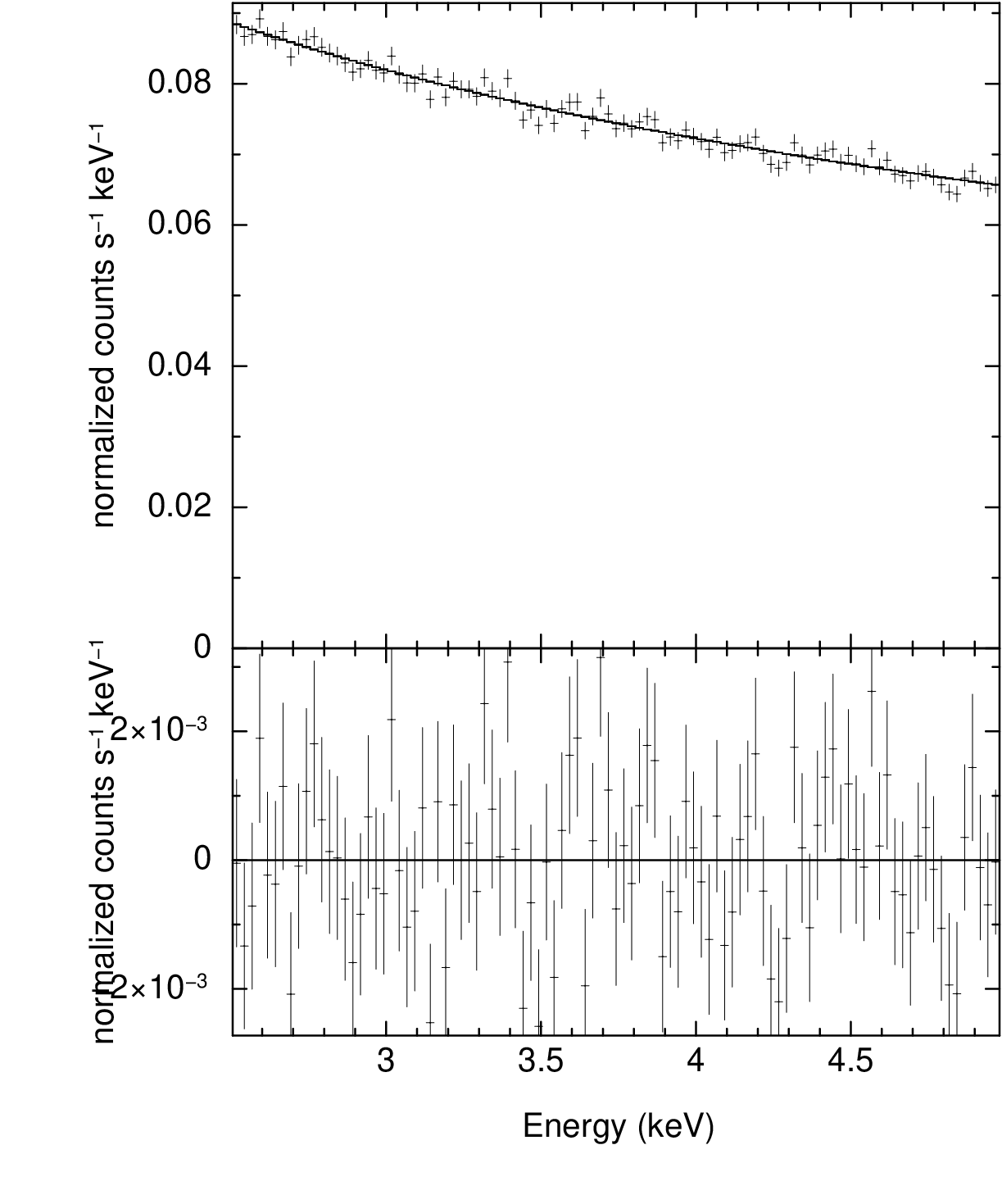}\qquad
	\includegraphics[width=0.93\columnwidth]{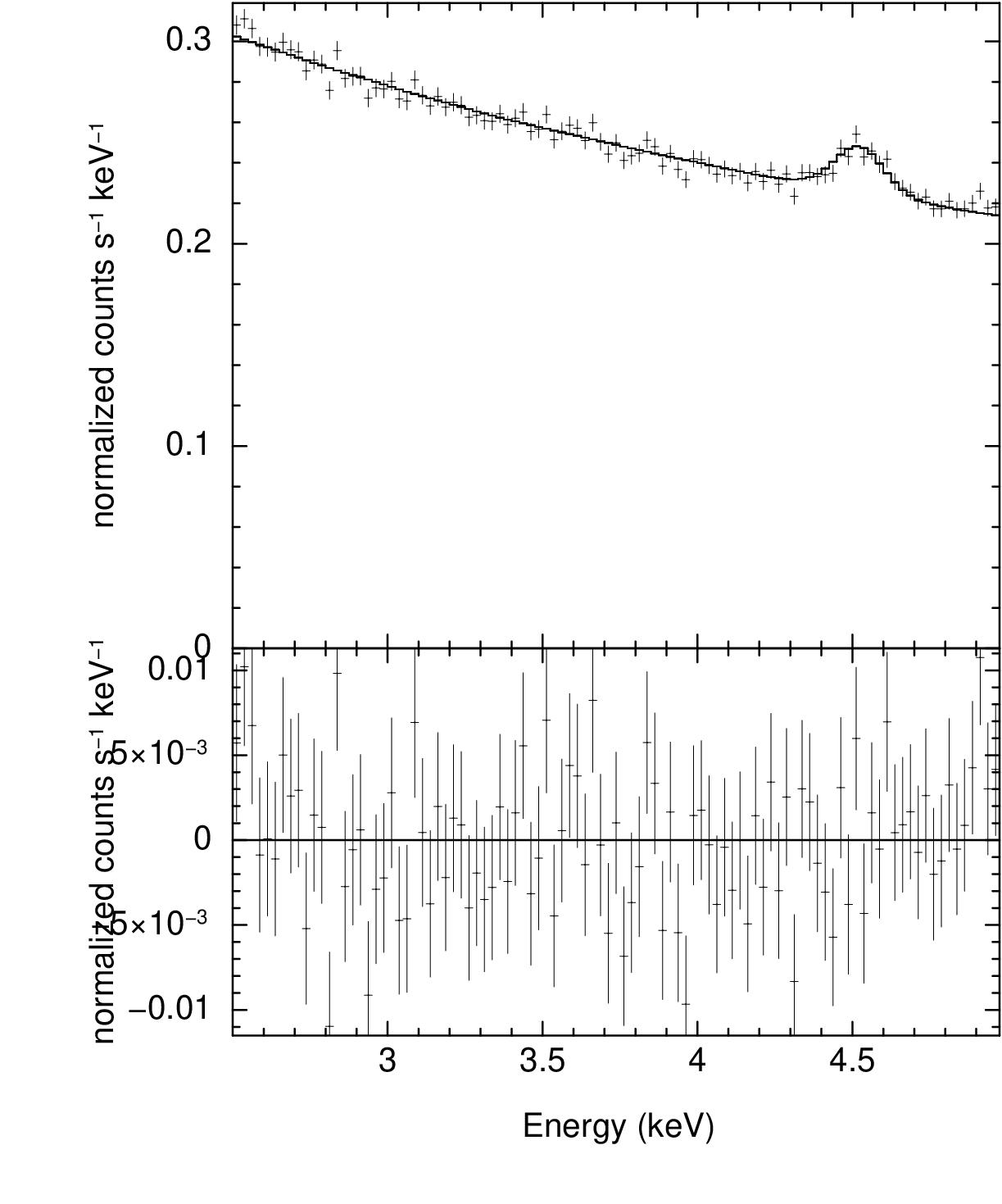}
	\end{centering}
\caption{\small {\em Left}:  Combined MOS spectrum and residuals in the 2.5-5.0 keV energy range fit to an unfolded, single power law.  For visual effect here the spectrum has been binned by a factor of five.  {\em Right:} Combined PN spectrum and residuals (with a similar factor of five binning) in the 2.5-5.0 keV energy range fit to an unfolded single power law plus an instrumental line due to Ti K$\alpha$ emission at 4.51 keV.  A weaker Ti K$\beta$ line can be seen at 4.93 keV but has no effect on the 3.5 keV line constraints.}
\label{fig:spec}
\end{figure*}

It is important to acknowledge that null results obtained so far are still compatible with a non-standard origin for the 3.5 keV line. Notably, axion-like particle conversion in magnetic fields \citep{Cicoli:2014bfa, Alvarez:2014gua} could reproduce the morphology of the 3.5 keV line in Perseus reported in \cite{Carlson:2014lla}; other possibilities include, for example, inelastic excited dark matter \citep{Finkbeiner:2014sja}. In all such instances, the signal strength scales non-trivially with the integrated dark matter mass along the line of sight, or it depends sensitively on astrophysical conditions such as the magnetic field strength.

\cite{Lovell:2014lea} used N-body simulations from the Aquarius project \citep{aquarius} to estimate the flux ratio for a standard dark matter decay process across different targets, including for the Draco dwarf spheroidal galaxy (dSph) and the Galactic center (GC). This ratio has a certain statistical distribution, which depends on the choice of the placement of the observer. The central finding of that study is that a 1.3 Ms long XMM-Newton observation of the Draco dSph would enable the discovery or exclusion at the 3$\sigma$ level of a dark matter decay interpretation of the 3.5 keV signal. 

Here, we utilize recent, deep archival XMM-Newton observations of the Draco dSph to test a dark matter decay origin for the 3.5 keV line. We find no evidence of a line in either the MOS or PN data, and we are able to rule out a dark matter decay origin at greater than the 99\% confidence level.

The remainder of this manuscript has the following structure: we describe the XMM observations and data reduction in the following section \ref{sec:data}; we then describe our flux calculation and compare with the flux limits from the XMM MOS and PN data in section \ref{sec:analysis}, and we present our conclusions in the final section \ref{sec:conclusions}.

\section{XMM Data Analysis}\label{sec:data}

Draco was observed by XMM-Newton in 31 separate observations, 5 in 2009 (PI Dhuga) and 26 in 2015 (PI Boyarsky), with individual exposure times ranging from 17 to 87 ksec and a total time in all observations of 1.66 Msec.  We reprocessed all 31 observations using standard procedures and utilizing the XMM SAS\footnote{http://xmm.esac.esa.int/sas/} and ESAS \citep{esas,esas2} software packages.  Starting from the Observation Data Files, the raw EPIC data was pipeline-processed with the {\tt emchain} and {\tt epchain} tasks.  Flare filtering was carried out with the ESAS tasks {\tt mos-filter} and {\tt pn-filter}; these time periods of increased particle background due to soft protons can lead to background levels elevated by two orders of magnitude and are thus removed from the data.  Unfortunately, in the case of Draco particle background flaring was significant in many of the observations.  For the two MOS detectors, two observations (ObsID 0603190401 and 0770190601) were almost entirely contaminated by flaring, and we removed these from our final data set; the other observations had reduced usable exposure times.  The net exposure time after filtering was a little over one Msec for each MOS detector with a total time for both detectors of 2.1 Msec.  The PN detector is typically more effected by particle flaring than the MOS detectors, and we found that only 20 of 31 observations had flares satisfactorily removed by {\tt pn-filter}; for these observations the net usable exposure time for PN was 0.58 Msec.

\begin{figure*}
\begin{centering}
	\includegraphics[width=0.97\columnwidth]{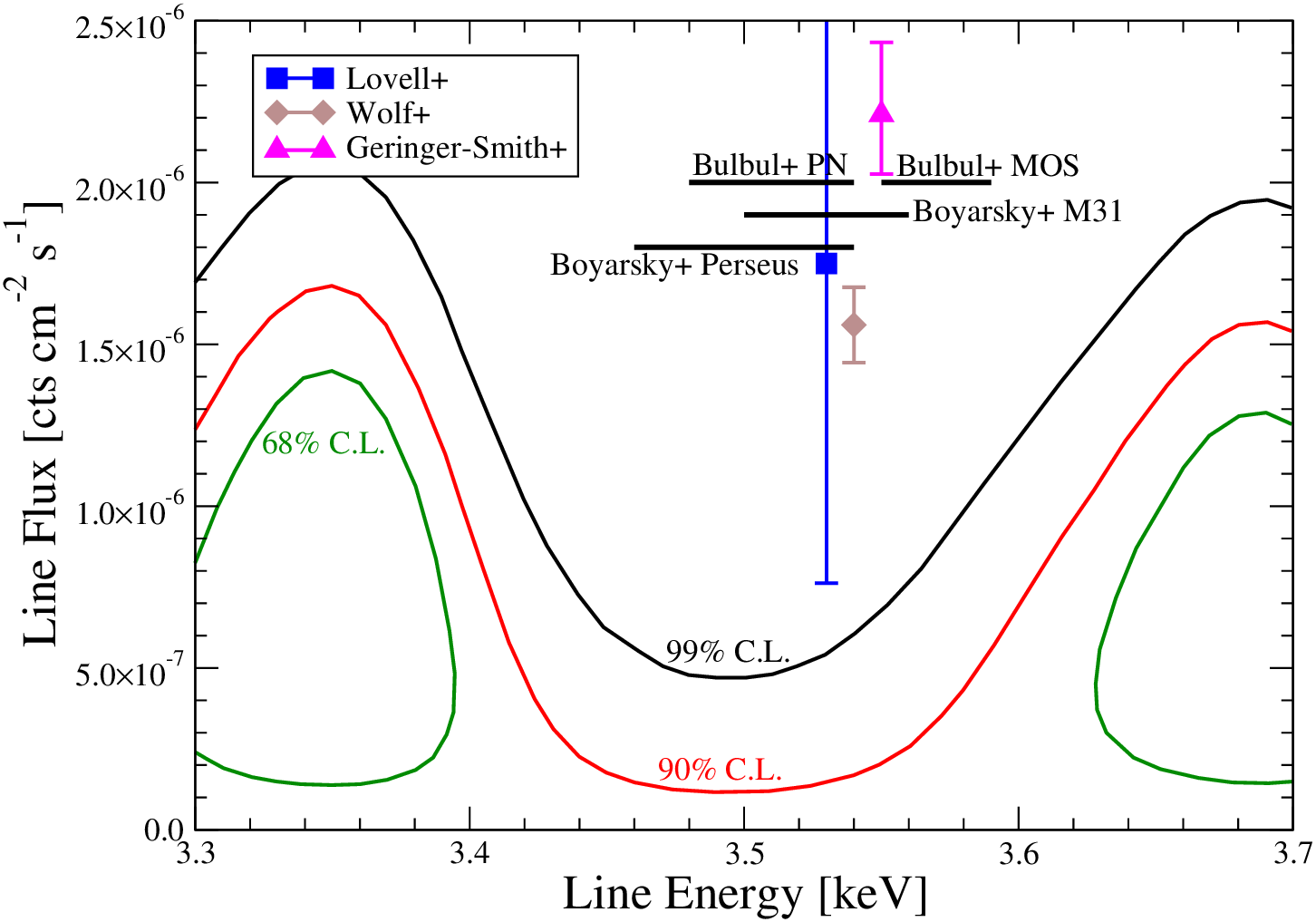}\qquad
	\includegraphics[width=0.97\columnwidth]{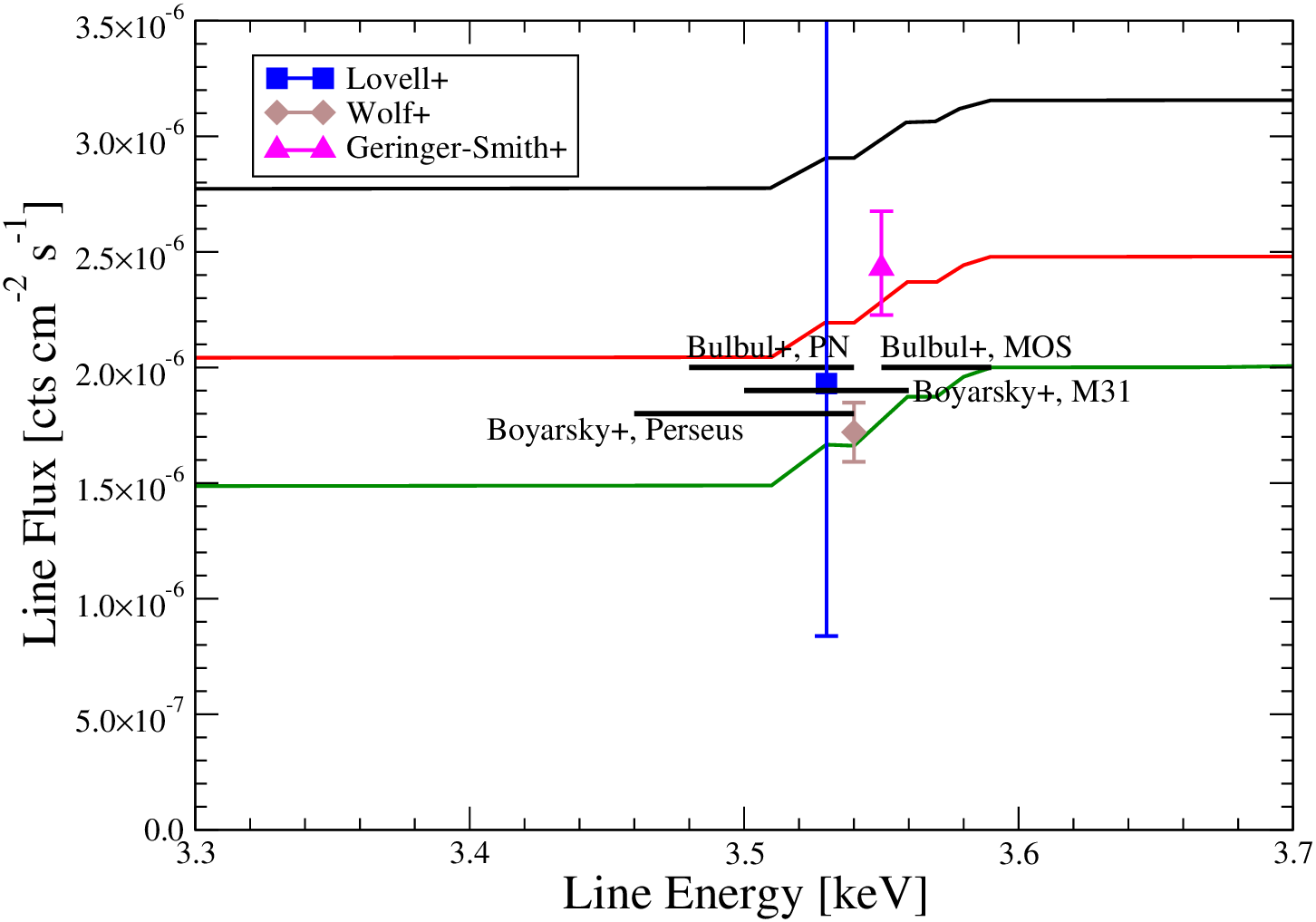}
	\end{centering}
\caption{\small {\em Left}:  Limits on the flux of a line in the energy range between 3.3 and 3.7 keV from MOS observations of the Draco dSph, at the 68\%, 90\% and 99\% C.L. (green, red and black lines, respectively) and predictions for the flux of a 3.5 keV line assuming a dark matter decay origin for the line detected at that energy from stacked clusters of galaxies and from the Milky Way center (see text for details). The horizontal black lines indicate the 1$\sigma$ energy range for the line position as inferred by Boyarsky et al. 2014 for Perseus ($3.50\pm0.04$ keV) and for M31 ($3.53\pm0.03$ keV) and by Bulbul et al. 2014 from cluster observations ($3.57\pm 0.02$ and $3.51\pm0.03$ keV for their ``full sample'' MOS and PN results, resepctively); {\em Right}: same, for PN observations (note the difference in vertical scale).}
\label{fig:flux}
\end{figure*}

Point sources were detected and removed separately from each observation using the ESAS task {\tt cheese}; point source detection was run on broad-band images (0.4-7.2 keV) with a flux limit of $10^{-14}$ erg cm$^{-2}$ s$^{-1}$ and a minimum separation of 10 arcsec.  Low exposure regions are likewise masked by {\tt cheese}.  Spectra were extracted from the full field-of-view from each detector in each flare-filtered observation; however, for the MOS1 detector CCDs 3 and 6 were excluded due to micrometeroid damage.  Spectra and corresponding redistribution matrix
files (RMF) and ancillary response files (ARF) for the 0.4-7.2 keV range were created using {\tt mos-spectra} and {\tt pn-spectra} in the ESAS package.  The individual spectra and response files were co-added using the routines {\tt mathpha},  {\tt addrmf}, and {\tt addarf} in the {\tt FTOOLS} package \citep{ftools}.  Combined RMF and ARF files were weighed by the relative contribution of each observation to the total exposure time.  The spectra and responses for the MOS1 and MOS2 cameras were combined in to a single summed MOS spectrum, while the spectra and responses for the PN detector were combined separately.

Spectral modeling employed the energy range between 2.5 keV and 5 keV.  This energy range was chosen to exclude strong instrumental emission lines while being much, much broader than the energy resolution of the detectors ($\sim 100$ eV).  At these energies, the X-ray background is dominated by the quiescent particle background \citep{esas2} which we model with an unfolded, power law (no vignetting) in {\tt XSPEC} \citep[version 12.8.1p, ][]{xspec}.  As shown in Fig.~\ref{fig:spec}, the combined MOS spectrum in the 2.5-5 keV range is well fit by an unfolded, single power law alone with reduced $\chi^2 = 0.96$ ($\chi^2=475/497$ degrees of freedom).  Adding a Gaussian line between 3.4 and 3.6 keV gives no improvement to the fit, and a line at these energies with a flux greater than $\sim 10^{-6}$ photons cm$^{-2}$ s$^{-1}$ is excluded, as shown in Fig.~\ref{fig:flux}.  The confidence contours in Fig.~\ref{fig:flux} are determined based on the change in the fit statistic when stepping over the relevant parameters using the {\tt steppar} command in {\tt XSPEC}.  The combined PN spectrum is well fit by an unfolded power law plus an instrumental line due to Ti K$\alpha$ emission (4.51 keV), which we model as a narrow Gaussian (Fig.~\ref{fig:spec}, right).  The reduced $\chi^2$ for this fit is 0.99 ($\chi^2=490/495$ degrees of freedom).  Again, adding a Gaussian line between 3.4 and 3.6 keV gives no significant improvement to the fit.  The fit is somewhat improved by adding a second instrumental line, Ti K$\beta$, at 4.93 keV, but this feature has no effect on the 3.5 keV line constraints.  As can be seen from Fig.~\ref{fig:flux}, right the upper limit on the flux of a line near 3.5 keV from the PN data is weaker than from the MOS data given the shorter usable exposure time but does serve as additional confirmation of the lack of a 3.5 keV line from Draco.

\section{Flux Limits and Constraints on Dark Matter Decay}\label{sec:analysis}
We utilize three distinct predictions for the 3.5 keV line flux that should have been observed with the XMM observations described above for a dark matter decay origin . The first one makes use of the results of \cite{Lovell:2014lea}, which calculated the flux expected from a 14 arcmin angular region around Draco given the brightness of the 3.5 keV line observed from the Galactic Center and the ratio of the flux from Draco-like halos and from the Galactic center as extrapolated from the Aquarius simulation. The resulting distribution in predictions is bracketed by the range 
$$
F=(1.0-5.2)\times10^{-6}\ {\rm cts}\ {\rm cm}^{-2}\ {\rm s}^{-1},
$$
where the lower and upper values bracket 95\% of the predictions, and with the most-probable value being $F=2.3\times 10^{-6}\ {\rm cts}\ {\rm cm}^{-2}\ {\rm s}^{-1}$ (see especially their Appendix C3 for additional details on assumptions and method). We calculated that the point source masking we adopt and the non-uniform coverage (e.g. from the lost MOS CCDs and chip gaps) described in the previous section suppress the predicted flux to 77\% of its un-masked value (we neglect the additional signal from the annulus between 14 and 15 arcmin). We calculated the fraction of masked signal employing the dark matter density profile and distance to Draco quoted in \cite{Abdo:2010ex}. We verified that the impact on the masking fraction of varying the parameters in the dark matter density profile and the distance within their 2-$\sigma$ range is in all cases smaller than 0.1\%. We show the resulting range of expected fluxes in Fig.~\ref{fig:flux} with the vertical blue bar, and indicate the central value with a full blue square.

We additionally considered two alternate predictions for the flux expected in the data we analyze for the decaying dark matter scenario. We followed the procedure outlined in \cite{Malyshev:2014xqa}. There, two alternate determinations of the mass within half-light radius were adopted to estimate the integrated line of sight dark matter column density for Draco, based on the results of the analyses of \cite{wolfetal} and \cite{geringer-smithetal}. \cite{Malyshev:2014xqa} then proceeded to add, for the direction of Draco, the flux from dark matter within the Milky Way, and added to it the prediction for the component from the Draco dSph itself. For the former, \cite{Malyshev:2014xqa} presents a ``mean'' flux based on the ``favored NFW'' Milky Way dark matter halo of \cite{Klypin:2001xu}, as well as a very conservative ``minimal'' model based on a ``maximal disk'' halo structure. We choose to show our predictions for the flux in Fig.~2 for the mean flux from the Milky Way, but it is a straightforward exercise (and one that does not impact our results or conclusions qualitatively) to rescale them for the minimal Milky Way model. The predicted flux was normalized, as in \cite{Malyshev:2014xqa} and in \cite{Lovell:2014lea}, to the parameters corresponding to the best-fit point of the cluster observations of \cite{Bulbul:2014sua}. We show the predictions for the Draco halo determination in  \cite{wolfetal} with brown diamonds and \cite{geringer-smithetal} with purple triangles. The uncertainties we show in the figure reflect the uncertainties from the determination of the halo parameters as in \cite{Malyshev:2014xqa}. As we did for the predictions from \cite{Lovell:2014lea}, we corrected the predicted flux quoted in \cite{Malyshev:2014xqa} for masking in our observations, and for the larger, in this case, angular region we utilize compared to the flux predictions in \cite{Malyshev:2014xqa}. 
We illustrate our results in Fig.~\ref{fig:flux}. We indicate with green, red and black lines the 68\%, 90\% and 99\% Confidence Level (C.L.) limits on the maximal allowed flux associated with a line at the energy indicated by the x-axis. We also show the predictions for the line flux described above as well as the range of energies for the line reported from cluster observations as described in \cite{Bulbul:2014sua} and the range obtained by \cite{Boyarsky:2014jta} from observations of the Perseus cluster and of M31.

As can be seen in Fig.~\ref{fig:flux}, the lack of a detected line in the MOS data rule out at higher than 99\% confidence level a line with even the most conservative predicted fluxes based on a conservative range of possible density profiles for the Draco dwarf.  Specifically, the lower limit on the predicted Draco line flux of \cite{Lovell:2014lea} based on the brightness of the line observed in the Galactic Center is excluded at 99.1\%; the lower limit on the predicted flux given the observed stacked cluster line flux are excluded at better than 99.999\% for either the \cite{wolfetal} or the \cite{geringer-smithetal} profiles.  Therefore, a generic dark matter decay origin of the 3.5 keV line feature is highly unlikely.  

\begin{figure}
\begin{centering}
	\includegraphics[width=0.9\columnwidth]{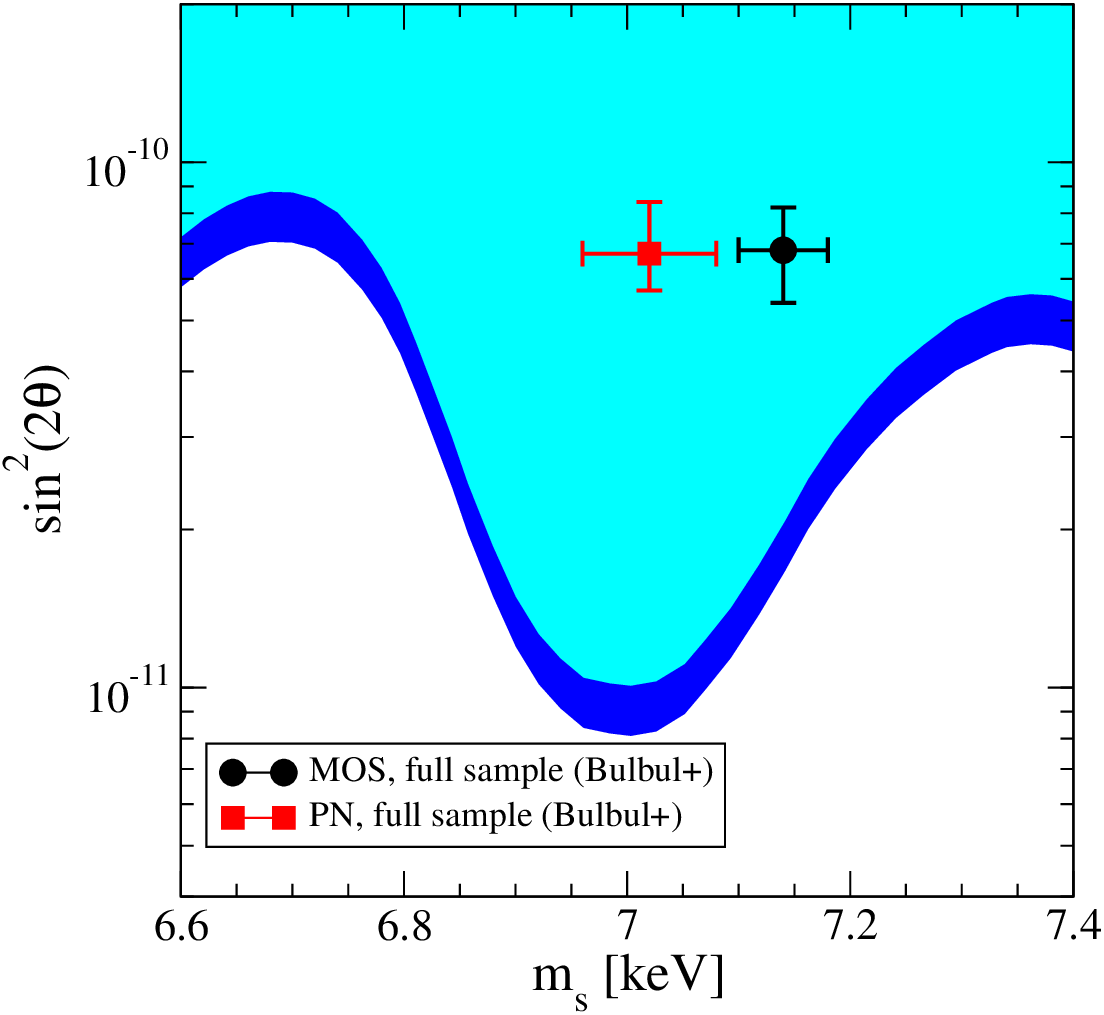}
		\end{centering}
\caption{\small Constraints on the parameter space of sterile neutrinos, defined by the particle's mass $m_s$ and mixing angle with active neutrinos, $\theta$. The cyan-shaded region is excluded, at the 2$\sigma$ level ($\sim$95\% C.L.), by Draco MOS observations, using the most conservative Milky Way dark matter density profile considered in Malyshev et al. (2014), while the blue-shaded region employs the nominal ``favored NFW'' profile, which we also use for Fig.~\ref{fig:flux}.}
\label{fig:limits}
\end{figure}

In Fig.~\ref{fig:limits}, we show constraints on the sterile neutrino parameter space in terms of the particle's mass $m_s$ and mixing angle with active neutrinos $\theta$ given the line flux limits from the MOS Draco observations, in the relevant mass range for a dark matter decay interpretation of the 3.5 keV line. The cyan shaded region is excluded at the 2$\sigma$ level, and assumes the central value for the \cite{geringer-smithetal} dark matter halo parameters for Draco, and the most conservative Milky Way halo considered in \cite{Malyshev:2014xqa} (corresponding to the ``maximal disk model'' of \cite{Klypin:2001xu}). The blue shaded region, instead, adopts the default ``favored NFW'' Milky Way dark matter halo density profile \citep{Klypin:2001xu}.  

Taking the most conservative possible assumptions both for the flux from Draco and from the Milky Way Galactic halo (corresponding to the predictions of \cite{wolfetal} for the flux from Draco and the most conservative Milky Way halo of \cite{Klypin:2001xu}), we are able to set a lower limit on the lifetime of a 7 keV sterile neutrino decaying into a 3.5 keV line of $\tau>2.7\times 10^{28}$ s (95\% C.L.), corresponding to a mixing angle $\sin^2(2\theta)<1.6\times 10^{-11}$ (95\% C.L.). Our most conservative limits are thus more than a factor 4 below the favored mixing angle predicated by a dark matter decay interpretation of the 3.5 keV line signal \citep[$\sin^2(2\theta)\approx7\times 10^{-11}$, ][]{Bulbul:2014sua}. 

After our paper appeared, \cite{ruchayskiy15} analyzed essentially the same data set as this work.  They come to a similar conclusion that no 3.5 keV line is detected in Draco, though their limit on the flux of the line is less stringent than ours.  The main difference appears to stem from the fact that they find a mild excess in the PN data which we do not see.  Here we comment briefly on some of the differences in the two analyses.  \cite{ruchayskiy15} make three comments on our analysis. 1) We do not include an extragalactic background power law in addition to the particle background; 2) the $\sim 1 \sigma$ excesses in the MOS spectrum near 3.35 keV and 3.7 keV might be due to un-modeled weak instrumental lines (from K K$\alpha$ and Ca K$\alpha$ at 3.31 and 3.69 keV); 3) we do not jointly fit the MOS and PN data.  We have now tried all three of these variations and find that they all have a negligible effect on our results.  Adding additional lines or background components as in points 1) and 2) is not warranted by the data nor does it significantly improve our fits when done.  In fact, we find that when we do add these components our flux limit on a line near 3.5 keV in the MOS data is actually {\em strengthened}, albeit slightly, lowering the flux limit by 10\%.  We note that when adding an additional power law for the extragalactic background in addition to the unfolded power law for the particle background, we expanded the energy range to 2.5-7 keV and added lines for the strong Cr, Mn, and Fe instrumental lines in this range, but again we find no excess near 3.5 keV.  Here the power law components are free to vary; the best fit photon indices are 1.5 for the extragalactic power law and 0.33 for the unfolded power law with a reduced $\chi^2$ of 0.99 ($\chi^2=876/885$ degrees of freedom).

Results from combined fits to the MOS and PN data must be interpreted with care as there are known offsets in the relative calibration of the two camera types.  In particular, fits to a stack of bright sources show an offset of 5-8\% between MOS and PN \citep{read14} above 3 keV, which is similar to or larger than the ratio of the predicted 3.5 keV line flux to the neighboring continuum.  In performing a joint fit to the MOS and PN spectra, we separately fit the continuum models for the two instruments, but jointly fit the 3.5 keV line energy and flux.  We again find no excess near 3.5 keV.  Our limit on the flux of a line near 3.5 keV is weakened only slightly (by 25-30\%) from the MOS-only limit shown in the left panel of Fig.~\ref{fig:flux}, and we still exclude the most conservative predictions from \cite{Lovell:2014lea} at the 99\% confidence level.

The primary reason that \cite{ruchayskiy15} quote a weaker limit on the flux of a line from Draco appears to be due to the fact that they find a $\sim 2 \sigma$ excess in the PN which is not present in our analysis.  It is unclear why they find an excess where we do not, but we note that in their spectral fits they include data up to 10 keV where the effective area is both rapidly dropping and a factor of $\sim 4$ lower than at 3 keV for PN, and more than an order of magnitude lower for MOS.  In addition, their background model, given the large energy range, includes 13 line components and two power laws compared to our single power law.

\section{Discussion and Conclusions}\label{sec:conclusions}

Using $\sim1.6$ Msec observations of the Draco dSph with XMM-Newton we were able to obtain one of the most stringent constraints on a dark matter decay origin for the 3.5 keV line observed from clusters of galaxies and from the Milky Way center. Our results rule out a dark matter decay interpretation with greater than 99\% C.L., and, under very conservative assumptions on the relevant dark matter density profiles, imply a lower limit on the dark matter lifetime of $\tau>2.7\times 10^{28}$ s at 95\% C.L. for a dark matter mass of 7 keV radiatively decaying to a two-body final state with one photon.

In view of the results presented here, and in view of the recent re-assessment of the potassium abundance \citep{Phillips:2015wla}, we conclude that the most probable counterpart to the 3.5 keV line observed towards the Milky Way center and from individual and stacked observations of clusters of galaxies are atomic de-excitation lines of the K XVIII ion. Charge-exchange processes might also provide an alternate astrophysical explanation \citep{Gu:2015gqm}. Scenarios advocating new physics where a 3.5 keV signal is suppressed in dwarf galaxies, such as an axion-like particle conversion to 3.5 keV photons in the presence of a magnetic field, are not ruled out unless Occam's razor is advocated. 

Future observations of clusters and of the Galactic center with Astro-H remain a priority to pinpoint the physical origin and the nature of the 3.5 keV line, while, in view of our results, additional deep observations of local dwarf galaxies with current or future telescopes are unlikely to advance our understanding of this particular feature.

\section*{Acknowledgments}
%
\noindent We would like to thank the referee for their valuable comments on our paper.  TJ is partly supported by NSF AST 1517545. SP is partly supported by the US Department of Energy, Contract DE-SC0010107-001. 
%

\bibliography{galcenter}
\end{document}